\documentclass[twocolumn,fleqn,final,5p,times]{elsarticle}
\usepackage[latin9]{inputenc}
\usepackage{color}
\usepackage{amstext}
\usepackage{graphicx}
\makeatletter
\usepackage{lineno}
\modulolinenumbers[5]
\journal{Phys. Lett. A}

\biboptions{numbers,sort&compress}
\makeatother

\begin{document}
\begin{frontmatter}

\title{Scattering coefficients and bound states for high-energy transparent
$\delta-\delta^{\prime}$ interactions}

\author[ufpa]{Danilo C. Pedrelli\corref{cor1}}

\ead{danilo.pedrelli@icen.ufpa.br}

\author[ufpa]{Jeferson Danilo L. Silva}

%\ead{jeferson.silva@icen.ufpa.br}

\author[ufpa]{Alessandra N. Braga}

%\ead{alessandranb@ufpa.br}

\author[ufpa]{Danilo T. Alves}

%\ead{danilo@ufpa.br}

\cortext[cor1]{Corresponding author}

\address[ufpa]{Faculdade de Física, Universidade Federal do Pará, Belém,
Pará, Brazil}
\begin{abstract}
We propose a model for energy-dependent  $\delta-\delta^{\prime}$ interactions
which yields scattering coefficients exhibiting full transmission for high-energy incident particles, also computing the bound solutions
in one-dimension nonrelativistic quantum mechanics.
\end{abstract}
\end{frontmatter}

%%%%%%%%%%%%%%%%%%%%%%%%%%%%%%%%%%%%%%%%%%%%%%%%%%%%%%%%%%%%%%%%%%%%%%%%%%%%%%%%%%%%%
\section{Introduction}

Point interactions play an important role in the class of solvable
quantum mechanical models representing systems with short range, but
strong, potentials. The first remarkable investigations on these kind of interactions
were done by Kronig and Penney \cite{Kronig-Penney-31}, Bethe and Peierls \cite{Bethe-Peierls-35}
and Thomas \cite{Thomas-35}, with fundamental
impact on the fields of condensed matter and nuclear physics. Several
fundamental mathematical aspects of point interactions have been studied
by Berezin and Faddeev \cite{Berezin-61}. Albeverio \emph{et al.}
\cite{Albeverio-1,Albeverio-2,Albeverio-3,Albeverio-98} obtained
a family of parameters that describe general solutions for point
interactions in the context of nonrelativistic quantum mechanics, requiring the Hamiltonian
to be self-adjoint and removing the interaction
point from the space of the corresponding free Hamiltonian. Šeba \emph{et al.} \cite{Seba-86,Seba-96,Seba-00}
and other authors \cite{Coutinho-97,Patil-94,Exner-95} have worked
in a similar way, constructing general models for one or more point
interactions in different dimensions.

A pure $\delta$ distribution, which is the simplest case of a point
interaction, is considered in many textbooks (see, for instance, \cite{Cohen}) as follows
(hereafter $m=\hbar=1$):
\begin{equation}
H=\frac{1}{2}\frac{d^{2}}{dx^{2}}+\mu\delta(x),\quad\mu\in\mathbb{R}.
\label{pure-delta}
\end{equation}
For $\mu>0$ the potential is repulsive, leading to the scattering
solution. For a monochromatic wave coming from the left we have: $\psi\left(x\right)=s(k)e^{ikx}$,
for $x>0$, and $\psi(x)=e^{ikx}+r(k)e^{-ikx}$, for $x<0$ (with
$k=\sqrt{2E}$ and $E>0$). Computation of the reflection $r\left(k\right)$
and transmission $s\left(k\right)$ coefficients furnishes \cite{Cohen}:
\begin{equation}
r(k)=-\frac{\mu}{\mu+ik},\quad k=\sqrt{2E},
\end{equation}
and
\begin{equation}
s(k)=\frac{ik}{\mu+ik},\label{transm0}
\end{equation}
what shows that, for Dirac delta potentials, the scattering coefficients are naturally energy-dependent and lead to full transmission in the limit of high-energy incident particles,
\begin{equation}
\lim_{k\rightarrow\infty}s(k)=1.\label{full0}
\end{equation}
Whether for $\mu<0$ the potential is attractive, also allowing bound solutions. Since the potential
is zero everywhere except at the origin, we expect that the bound
state solution vanishes for $x\rightarrow\pm\infty$, what enables
us to write the wave function as $\psi(x)=Ae^{-\kappa|x|}$ ($A\in\mathbb{R}$
and $\kappa=\sqrt{-2E}$). Calculation of the energy gives just one
solution, namely \cite{Cohen}
\begin{equation}
E=-\mu^{2}/2<0.\label{energy0}
\end{equation}

For a pure delta distribution, the wave function is continuous everywhere. On the other hand, potentials involving the
derivative of a delta function can generate discontinuity \cite{Albeverio-1}.
A few regularization methods have been developed in order to describe
quantum mechanics with such potentials \cite{Coutinho-97,Zolotaryuk-03,Zolotaryuk-06,Zolotaryuk-08,Nogami-07},
however, those methods do not give the same results for the transmission
coefficient. Kurasov \cite{Kurasov-96} suggested that, to properly
define self-adjoint operators in this case, it is necessary to use
distribution theory for discontinuous functions, deriving the boundary
conditions at the point where the interaction occurs. Taking this
into account, Gadella \emph{et al.} \cite{Gadella-2009} have determined
the bound state and scattering coefficients for such interactions
in the context of the distribution theory proposed by Kurasov, obtaining
results without making use of any regularization scheme. They have
investigated the Hamiltonian \cite{Gadella-2009} 
\begin{equation}
H=\frac{1}{2}\frac{d^{2}}{dx^{2}}-\mu\delta(x)+\lambda\delta^{\prime}(x),\quad\mu\in\mathbb{R}_{+}^{*},\,\lambda\in\mathbb{R},\label{primo}
\end{equation}
where $\mu$ is made positive in order to ensure the existence of
bound solutions. The scattering coefficients for a left-incident
monochromatic wave are \cite{Gadella-2009} 
\begin{equation}
r(k)=-\frac{\mu+2ik\lambda}{\mu+ik\left(1+\lambda^{2}\right)},\quad k=\sqrt{2E},
\end{equation}
and
\begin{equation}
s(k)=\frac{ik(1-\lambda^{2})}{\mu+ik(1+\lambda^{2})}.\label{eq:hhh}
\end{equation}
As mentioned before, $\delta^{\prime}$ interactions generate discontinuous
wave functions and, hence, the bound state solution at the right side
of the interaction has a different amplitude compared to the left
one, namely: $\psi(x)=Ae^{-\kappa x}\Theta(x)+Be^{\kappa x}\Theta(-x)$, where
$A,\,B\in\mathbb{R}$, $\kappa=\sqrt{-2E}$, and $\Theta(x)$ is the Heaviside function. The bound state energy
is \cite{Gadella-2009}:
\begin{equation}
E=-\frac{\mu^{2}/2}{(1+\lambda^{2})^{2}} \label{gadeq}.
\end{equation}
One can observe that the parameter $\lambda$ decreases the energy
amplitude in comparison to the bound energy of a pure $\delta$ interaction,
Eq. (\ref{energy0}). As well as before, there is only one bound state
solution.

In the limit of high energies, Eq. (\ref{eq:hhh}) becomes
\begin{equation}
\lim_{k\rightarrow\infty}s(k)=\frac{1-\lambda^{2}}{1+\lambda^{2}}<1,\quad\lambda\neq0.\label{eq:full1}
\end{equation}
In contrast to the pure $\delta$ case [Eq. (\ref{full0})], the above result is in disagreement with the physical intuition that
every incident wave with infinite energy should totally pass through
any obstacle. This misbehavior was also noticed in Ref. \cite{Zolotaryuk-03}. In the context of quantum field theory, Braga \emph{et al.} \cite{Braga} have also noticed that mirrors simulated by $\delta-\delta^{\prime}$ potentials \cite{Munoz-Castaneda-01,Munoz-Castaneda-2}
are not completely transparent in the limit of high-frequencies and,
to solve this problem, they considered that the coupling parameters
of the point interaction are frequency-dependent functions that vanish
for high frequencies. The correspondent procedure in quantum mechanics is to consider energy-dependent point interactions, what has been investigated by Coutinho \emph{et al.} \cite{Coutinho-JPA}, who define a set of boundary conditions that generates
energy-dependent point interactions.

In the present paper, we investigate a model for $\delta-\delta'$
interactions with an energy dependence in the way of Ref. \cite{Coutinho-JPA},
but restrict the parameters that define the interaction to be functions
that provide full transmission at high energies.
Straightforwardly, we compute the scattering coefficients and the
bound state, observing that energy-dependent potentials have the feature
of modifying the probability density, as explained in Ref. \cite{Lombard-04}.

This Letter is organized as follows. In Sec. \ref{sec:model} we discuss the $\delta-\delta^{\prime}$
interaction with energy-dependent coupling parameters, taking as basis  the model found in Ref. \cite{Coutinho-JPA}. In Sec. \ref{sec:Scattering-Coefficients}
we find the scattering coefficients and show that the energy-dependent
parameters lead to full transmission in the limit of high energies.
In Sec. \ref{sec:Bound-States} we determine the bound state and its
energy, pointing out some aspects of the model. The final remarks
are presented in Sec. \ref{sec:Final-Remarks}.

%%%%%%%%%%%%%%%%%%%%%%%%%%%%%%%%%%%%%%%%%%%%%%%%%%%%%%%%%%%%%%%%%%%%%%%%%%%%%%%%%%%%%
\section{The $\delta-\delta^{\prime}$ interaction with energy-dependent parameters\label{sec:model}}

We begin by shortly discussing the model for energy-dependent point
interactions developed in Ref. \cite{Coutinho-JPA}.

As outlined in Ref. \cite{Albeverio-1}, point interactions can be characterized
by the boundary conditions
\begin{equation}
\left(\begin{array}{c}
\psi_{+}^{\prime}\\
\psi_{+}
\end{array}\right)=U\left(\begin{array}{c}
\psi_{-}^{\prime}\\
\psi_{-}
\end{array}\right),\quad U=e^{i\theta}\left(\begin{array}{cc}
\alpha & \beta\\
\delta & \gamma
\end{array}\right),\label{eq:u}
\end{equation}
with  $\alpha$, $\beta$, $\delta$, $\gamma$
$\in\mathbb{R}$,
\begin{equation}
\alpha\gamma-\beta\delta=1,
\label{vinculo}
\end{equation}
and $\psi_{+}$ and $\psi_{-}$ being the right and left limits of the wave function at $x=0$, and $\psi_{+}^{\prime}$ and $\psi_{-}^{\prime}$ the same limits taken on its derivative. The parameter $\theta$ is some phase which is not taken
under consideration for the stationary states we are going to treat
here (for more details on the implications of such phase see Ref. \cite{Coutinho-JPA}),
so that in the present paper we make $e^{i\theta}=-1$.

For instance, let us consider the potential $V(x)=2c_{0}\delta(x)$,
which can be represented by the boundary conditions 
\begin{eqnarray}
\psi_{+}^{\prime}-\psi_{-}^{\prime} & = & c_{0}\left(\psi_{+}+\psi_{-}\right),\quad c_{0}\in\mathbb{R},\label{bound1}\\
\psi_{+}-\psi_{-} & = & 0,\label{bound2}
\end{eqnarray}
from which one concludes, considering Eq. (\ref{eq:u}), that
\begin{equation}
U=\left(\begin{array}{cc}
1 & 2c_{0}\\
0 & 1
\end{array}\right).
\end{equation}
In order to include an energy dependence on the parameters of the
interaction, Coutinho \emph{et al.} \cite{Coutinho-JPA} imposed
a dependence between $\psi_{\pm}^{\prime}$ and $\psi_{\pm}^{\prime\prime}$,
replacing Eq. (\ref{bound1}) by 
\begin{equation}
\psi_{+}^{\prime}-\psi_{-}^{\prime}=-c_{1}\left(\psi_{+}^{\prime\prime}+\psi_{-}^{\prime\prime}\right),\quad c_{1}\in\mathbb{R}.\label{bound3}
\end{equation}
From the Schrödinger equation for a stationary state with energy $E$,
\begin{equation}
\psi^{\prime\prime}=-E\psi\quad(x\neq0),\label{bound4}
\end{equation}
and Eqs. (\ref{bound2}) and (\ref{bound3}), it is possible to write
the energy-dependent boundary conditions by means of Eq. (\ref{eq:u}), with the matrix
\begin{equation}
U=\left(\begin{array}{cc}
1 & 2c_{1}E\\
0 & 1
\end{array}\right).
\end{equation}

An extension of the above example can be made by assuming 
\begin{equation}
c(E)=\sum_{n=0}^{\infty}c_{n}E^{n},\quad c_{n}\in\mathbb{R}.\label{c(E)}
\end{equation}
Hence, one can rewrite Eq. (\ref{bound1}) as
\begin{equation}
\psi_{+}^{\prime}-\psi_{-}^{\prime}=\sum_{n=0}^{\infty}\left(-1\right){}^{n}c_{n}\left[\psi_{+}^{\left(2n\right)}+\psi_{-}^{\left(2n\right)}\right],
\end{equation}
where $\psi^{(2n)}=d^{2n}\psi/dx^{2n}$. Within the above considerations,
the matrix $U$ becomes
\begin{equation}
U=\left(\begin{array}{cc}
1 & 2c(E)\\
0 & 1
\end{array}\right).
\end{equation}

Considering Eq. (\ref{vinculo}), the boundary conditions (\ref{eq:u}) can be written as
\begin{equation}
\psi_{+}^{\prime}-\psi_{-}^{\prime}=\xi_{1}\left(\psi_{+}+\psi_{-}\right)-\xi_{2}\left(\psi_{+}^{\prime}+\psi_{-}^{\prime}\right),\label{bound5}
\end{equation}
\begin{equation}
\psi_{+}-\psi_{-}=\xi_{2}\left(\psi_{+}+\psi_{-}\right)-\xi_{3}\left(\psi_{+}^{\prime}+\psi_{-}^{\prime}\right),\label{bound6}
\end{equation}
where $\xi_{1}$, $\xi_{2}$, $\xi_{3}$ $\in\mathbb{R}$. Straightforwardly,
from Eq. (\ref{eq:u}) follows that $U$ turns out to be
\begin{equation}
U=-\frac{1}{\Delta}\left(\begin{array}{cc}
\Delta-2\left(1-\xi_{2}\right) & -2\xi_{1}\\
2\xi_{3} & \Delta-2\left(1+\xi_{2}\right)
\end{array}\right),\label{boundary1}
\end{equation}
where
\begin{equation}
\Delta=\left(1+\xi_{2}\right)\left(1-\xi_{2}\right)+\xi_{2}\xi_{3},\quad\Delta\neq0.
\end{equation}
The parameters $\xi_{1}$, $\xi_{2}$, $\xi_{3}$ and $\Delta$ are
related to $\alpha$, $\beta$, $\delta$ and $\gamma$ by
\begin{equation}
\xi_{1}=-\frac{\beta\Delta}{2},\,\xi_{2}=\frac{1}{4}\left(\alpha-\gamma\right)\Delta,\,\xi_{3}=\frac{\delta\Delta}{2},\,\Delta=\frac{4}{2-\alpha-\gamma}.\label{param}
\end{equation}

Similarly to Eq. (\ref{c(E)}), it is possible to include an energy dependence
on these parameters by making \cite{Coutinho-JPA}
\begin{equation}
\xi_{j}\left(E\right)=\sum_{n=0}^{\infty}d_{jn}E^{n},\quad d_{jn}\in\mathbb{R},
\end{equation}
where $j=1,\,2,\,3$. Then, one replaces the following equations in Eqs. (\ref{bound5}) and
(\ref{bound6}):
\begin{equation}
\xi_{j}\left(\psi_{+}^{\prime}+\psi_{-}^{\prime}\right)\rightarrow\sum_{n=0}^{\infty}\left(-1\right)^{n}d_{jn}\xi_{j}\left[\psi_{+}^{\left(2n+1\right)}+\psi_{-}^{\left(2n+1\right)}\right],
\end{equation}
\begin{equation}
\xi_{j}\left(\psi_{+}+\psi_{-}\right)\rightarrow\sum_{n=0}^{\infty}\left(-1\right)^{n}d_{jn}\xi_{j}\left[\psi_{+}^{\left(2n\right)}+\psi_{-}^{\left(2n\right)}\right].
\end{equation}
Up to here, we have outlined the fundamental aspects of the model developed in Ref.
\cite{Coutinho-JPA}, which is applicable to general energy-dependent point interactions.
Taking this as basis to the construction of our model,
we consider a $\delta-\delta^{\prime}$ interaction with an energy-dependent potential,
intending to obtain full transmission at high energies. We determine the correspondent coupling parameters 
by the distribution theory for discontinuous functions
given in Ref. \cite{Kurasov-96}. In this sense, we propose a modified Hamiltonian which leads to 
the Schrödinger equation
\begin{equation}
\left[-\frac{1}{2}\frac{\partial^{2}}{\partial x^{2}}-\mu\delta(x)+
\hat{F}\delta^{\prime}(x)\right]\Psi\left(x,t\right)=i\frac{\partial}{\partial t}\Psi\left(x,t\right),
\label{nosso-modelo}
\end{equation}
where 
%$\lambda_{0}\in\mathbb{R}$, 
$\mu\in\mathbb{R}_{+}^{*}$ and $\hat{F}$ is the following operator
\begin{equation}
\hat{F}=\sum_{n=0}^{\infty}F_{n}\left(i\frac{\partial}{\partial t}\right)^{n},
\label{def-F}
\end{equation}
with the coefficients $F_{n}$ chosen so that, 
when setting $\Psi\left(x,t\right)=e^{-iEt}\psi(x)$, we obtain a
time-independent Schrödinger equation with an energy-dependent potential,
\begin{equation}
\left[-\frac{1}{2}\frac{d^{2}}{dx^{2}}-\mu\delta(x)+F(E)\delta^{\prime}(x)\right]\psi(x)=E\psi(x),\label{Hamil}
\end{equation}
with
\begin{equation}
\lim_{E\rightarrow\infty}F(E)=0,
\label{lim-F}
\end{equation}
in a way that the full transmission in the limit
of high-energy incident particles can be achieved.
Notice that we have not considered any change in the coupling parameter of the $\delta$ term, since the pure $\delta$ interaction naturally leads to full transmission at high energies.

As an example of a function that vanishes for $E \rightarrow \infty$, we investigate the particular case for which $F(E)$, now relabeled as $\lambda _{E}$, is given by
\begin{equation}
F(E)=\lambda_{E}=\lambda_{0}\exp(-E/E_{0}), \quad \lambda_{0} \in \mathbb{R}, \;  E_{0}>0.
\label{f(E)}
\end{equation}
According to Ref.
\cite{Kurasov-96}, the above Hamiltonian leads to
\begin{equation}
U=\left(\begin{array}{cc}
\frac{-2\mu}{1-\lambda_{E}^{2}} & \frac{1-\lambda_{E}}{1+\lambda_{E}}\\
\frac{1+\lambda_{E}}{1-\lambda_{E}} & 0
\end{array}\right).\label{matchingC}
\end{equation}
A comparison between Eq (\ref{matchingC}) and Eq. (\ref{boundary1}),
also using Eq. (\ref{param}), enables us to relate the coupling
parameters $\mu$ and $\lambda_{E}$ with $\xi_{j}$ and $\Delta$,
as well as with $\alpha$, $\beta$, $\delta$ and $\gamma$, as follows:
\begin{equation}
\xi_{1}=-\mu\frac{\left(1+\lambda_{E}\right)}{\lambda_{E}^{2}},\enskip\xi_{2}=\frac{1}{2\lambda_{E}},\enskip\xi_{3}=0,\enskip\Delta=-\frac{(1-\lambda_{E}^{2})}{\lambda_{E}^{2}},
\end{equation}
and
\begin{equation}
\alpha=\frac{1-\lambda_{E}}{1+\lambda_{E}},\enskip\beta=\frac{-2\mu}{1-\lambda_{E}},\enskip\delta=0,\enskip\gamma=\frac{1+\lambda_{E}}{1-\lambda_{E}}.
\end{equation}

In Sec. \ref{sec:Scattering-Coefficients}, we will determine the scattering coefficients and, in Sec. \ref{sec:Bound-States},
the bound state solution and its energy.

%%%%%%%%%%%%%%%%%%%%%%%%%%%%%%%%%%%%%%%%%%%%%%%%%%%%%%%%%%%%%%%%%%%%%%%%%%%%%%%%%%%%%
\section{Scattering Coefficients\label{sec:Scattering-Coefficients}}

\begin{figure}[t]
\begin{centering}
\includegraphics[width=0.9\columnwidth]{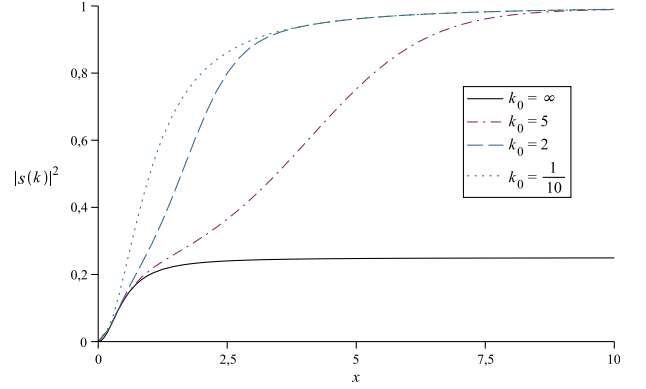} 
\par\end{centering}

\caption{Plot of the transmission coefficient $s(k)$ for several values of
the parameter $k{}_{0}$, $\mu=\lambda_{0}=1$.\label{fig:Plot-of-the}}
\end{figure}

The scattering solution of Eq. (\ref{Hamil}), for a left-incident monochromatic wave, is given by:
$\psi(x)=e^{ikx}+r(k)e^{-ikx}$, for $x<0$, and $\psi(x)=s(k)e^{ikx}$,
for $x>0$ (with $k=\sqrt{2E}>0$). Thus, from Eqs. (\ref{eq:u})
and (\ref{matchingC}), we obtain 
\begin{equation}
\left(\begin{array}{c}
iks\\
s
\end{array}\right)=\left(\begin{array}{cc}
\frac{-2\mu}{1-\lambda_{E}^{2}} & \frac{1-\lambda_{E}}{1+\lambda_{E}}\\
\frac{1+\lambda_{E}}{1-\lambda_{E}} & 0
\end{array}\right)\left(\begin{array}{c}
ik(1-r)\\
1+r
\end{array}\right),
\end{equation}
which provides the scattering coefficients 
\begin{equation}
r(k)=-\frac{\mu+2ik\lambda_{0}e^{-k^{2}/k_{0}^{2}}}{\mu+ik\left(1+\lambda_{0}^{2}e^{-2k^{2}/k_{0}^{2}}\right)},
\end{equation}
\begin{equation}
s(k)=\frac{ik\left(1-\lambda_{0}^{2}e^{-2k^{2}/k_{0}^{2}}\right)}{\mu+ik\left(1+\lambda_{0}^{2}e^{-2k^{2}/k_{0}^{2}}\right)}, \quad k_{0}=\sqrt{2E_{0}}.
\end{equation}
As expected, $\left|r(k)\right|^{2}+\left|s(k)\right|^{2}=1$. Finally,
we obtain that the transmission tends to one for $k\rightarrow\infty$,
\begin{equation}
\lim_{k\rightarrow\infty}s(k)=1.
\label{limite-ok}
\end{equation}

Figure \ref{fig:Plot-of-the} shows that for $k_{0}\rightarrow\infty$
(solid line), which recovers the model of Ref. \cite{Gadella-2009},
the transmission coefficient does not tend to one in the limit $k\rightarrow\infty$.
On the other side, we obtain full transmission at high energies for
any other value of the parameter $k_{0}$, showing that, as the energy
increases, the particle feels less and less the existence of the interaction.
With this result we show a way to manipulate the coupling parameter of the $\delta^{\prime}$ term in order to match the features
of a more realistic model.

%%%%%%%%%%%%%%%%%%%%%%%%%%%%%%%%%%%%%%%%%%%%%%%%%%%%%%%%%%%%%%%%%%%%%%%%%%%%%%%%%%%%%
\section{Bound States\label{sec:Bound-States}}

In the present section, using a procedure similar to that found in Ref. \cite{Gadella-2009},  which in turns is based on the distribution theory developed in Ref. \cite{Kurasov-96}, we obtain the bounded energy and wave function for our model.

Requiring that the bound state solution of Eq.
(\ref{Hamil}) vanishes for $x\rightarrow\pm\infty$, we
obtain 
\begin{equation}
\psi(x)=Ae^{\kappa x}\Theta(-x)+Be^{-\kappa x}\Theta(x),\quad\kappa=\sqrt{-2E},\label{BoundSt}
\end{equation}
where now $A$ and $B$ are energy-dependent parameters
(note that $A=\psi_{-}$ and $B=\psi_{+}$).

From the standard distribution theory for continuous wave functions, one gets the following equations \cite{Gadella-2009}:
\begin{equation}
\psi(x)\delta(x)=\psi(0)\delta(x),
\end{equation}
\begin{equation}
\psi(x)\delta^{\prime}(x)=\psi(0)\delta^{\prime}(x)-\psi^{\prime}(0)\delta(x).
\end{equation}
The extension to discontinuous wave functions can be made by using
the average approach \cite{Kurasov-96}, 
\begin{equation}
\psi(x)\delta(x)=\frac{\psi_{+}+\psi_{-}}{2}\delta(x),\label{average1}
\end{equation}
\begin{equation}
\psi(x)\delta^{\prime}(x)=\frac{\psi_{+}+\psi_{-}}{2}\delta^{\prime}(x)-\frac{\psi_{+}^{\prime}+\psi_{-}^{\prime}}{2}\delta(x),\label{average2}
\end{equation}
where it is understood that the wave function at the right and left
sides of the interaction point are not disjoint, and also that, for
$x>0$ or $x<0$, the features of continuous functions still hold,
e.g. the first and second derivative exist, with $\psi\left(x\right)$
and $\psi^{\prime\prime}(x)$ being square integrable functions \cite{Gadella-2009}.

\begin{figure}[t]
\begin{centering}
\includegraphics[width=0.9\columnwidth]{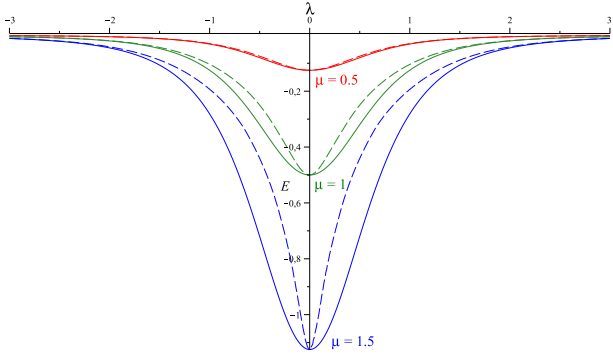} 
\par\end{centering}

\caption{Energy plot as function of $\lambda_{0}$ for $E_{0}=\infty$
(solid lines) and $E_{0}=1$ (dashed lines), and several values of $\mu$.\label{fig:energy}}
\end{figure}

By differentiating Eq. (\ref{BoundSt}) twice, we obtain 
\begin{equation}
\psi^{\prime\prime}(x)=\kappa^{2}\psi(x)-\kappa\left(A+B\right)\delta(x)+\left(B-A\right)\delta^{\prime}(x),\label{psi2}
\end{equation}
where we have used Eqs. (\ref{average1}) and (\ref{average2}). After
inserting Eq. (\ref{psi2}) into Eq. (\ref{Hamil}) we find 
\begin{equation}
\frac{\kappa(A+B)}{2}\delta(x)+\frac{(B-A)}{2}\delta^{\prime}(x)=\mu\psi(x)\delta(x)-\lambda_{E}\psi^{\prime}(x)\delta^{\prime}(x).\label{equality}
\end{equation}
Using the wave function given by Eq. (\ref{BoundSt}) together with
Eqs. (\ref{average1}) and (\ref{average2}), we obtain
\begin{equation}
\psi(x)\delta(x)=\frac{(A+B)}{2}\delta(x),\label{product1}
\end{equation}
\begin{equation}
\psi(x)\delta^{\prime}(x)=\frac{(A+B)}{2}\delta^{\prime}(x)-\frac{\kappa(A-B)}{2}\delta(x).\label{product2}
\end{equation}
Therefore, using (\ref{equality}), (\ref{product1}) and (\ref{product2}), we obtain 
\begin{equation}
(\kappa-\mu)(A+B)-(A-B)\kappa\lambda_{E}=0,\label{relation}
\end{equation}
\begin{equation}
(A-B)+\lambda_{E}(A+B)=0\quad\Leftrightarrow\quad\lambda_{E}=\frac{A-B}{A+B}.\label{parameter}
\end{equation}
The solution for $\lambda_{E}$ in terms of $\kappa$ is
\begin{equation}
\lambda_{E}=\left[1-\frac{\left(A-B\right)^{2}}{\left(A+B\right)^{2}}\right]\kappa.\label{momento}
\end{equation}
From the boundary conditions represented by Eq. (\ref{matchingC}), we have 
\begin{equation}
A-B=-\frac{2A\lambda_{E}}{1-\lambda_{E}},\label{alpha}
\end{equation}
\begin{equation}
A+B=\frac{2A}{1-\lambda_{E}}.
\end{equation}
Substituting the above relations into Eq. (\ref{momento}), we finally
get 
\begin{equation}
\kappa=\frac{\mu}{1+\lambda_{E}^{2}}=\sqrt{-2E},
\end{equation}
or, in terms of the energy, 
\begin{equation}
E=\frac{-\mu^{2}/2}{\left(1+\lambda_{0}^{2}e^{-2E/E_{0}}\right)^{2}}.\label{eq:Energy}
\end{equation}
This transcendental equation has only one solution,
which is negative independently of the parameters $\mu$, $\lambda_{0}$
and $E_{0}$.
%Notice that energy-dependent
%point interactions lead to more general classes of bound state
%solutions. Specifically, considering the point interaction given in Eq. (\ref{Hamil}), we obtain Eq. (\ref{eq:Energy}), from which, after taking the limit $E_{0}\rightarrow\infty$, we recover the Eq. (\ref{gadeq}) found in Ref. \cite{Gadella-2009}.
%
Figure \ref{fig:energy} shows the energy as function
of $\lambda_{0}$ for three different values of $\mu$, namely, $\mu=0.5$,
$\mu=1$ and $\mu=1.5$. Observing the shape of the curves for $E_{0}\rightarrow\infty$
(which corresponds to the model discussed in Ref. \cite{Gadella-2009}) and
those for the energy-dependent model (\ref{Hamil}) and (\ref{f(E)}) with $E_{0}=1$, we see that
the dashed curves become narrower in comparison to the solid curves as $\mu$ increases, what means that the parameter $E_{0}$ is responsible for a reduction of the bound energy magnitude. This effect is amplified as $E_{0}$ decreases.

%
%\caption{Energy plot as function of $\lambda_{0}$ for: $E_{0}=\infty$
%(solid lines) and $E_{0}=1$ (dashed lines); $\mu=0.5$ (lowest amplitude
%curves), $\mu=1$ (intermediate amplitude curves) and $\mu=1.5$ (greatest
%amplitude curves).\label{fig:energy}}
%\end{figure}

Formánek \emph{et al.} \cite{Lombard-04} have shown that, for energy-dependent
potentials, the usual definition of probability density does not satisfy
the continuity equation and, in order to solve this problem, they proposed the following modification
in the probability density:
\begin{equation}
\rho(x)=\left|\psi(x)\right|^{2}\left[1-\frac{\partial V(x,E)}{\partial E}\right],
\end{equation}
where $V(x,E)$ is an energy-dependent potential and $\rho(x)$ is the probability density. 
Hence, to properly normalize the wave function it is necessary to
redefine the norm \cite{Lombard-04},
\begin{equation}
\int_{-\infty}^{\infty}\psi^{*}(x)\left[1-\frac{\partial V(x,E)}{\partial E}\right]\psi(x)dx=1.\label{eq:norm}
\end{equation}

From the above relation and the properties of the $\delta^{\prime}(x)$
distribution given by Eq. (\ref{average2}), we are able to normalize
the wave function as follows: 
\begin{equation}
\int_{-\infty}^{\infty}\psi^{*}(x)\psi(x)dx-\frac{\lambda_{E}}{E_{0}}\left[\bar{\psi}^{\prime*}(0)\bar{\psi}(0)+\bar{\psi}^{*}(0)\bar{\psi}^{\prime}(0)\right]=1.
\end{equation}
where $\bar{\psi}\left(0\right)=(\psi_{+}+\psi_{-})/2$ and $\bar{\psi}^{\prime}\left(0\right)=(\psi_{+}^{\prime}+\psi_{-}^{\prime})/2$.
Recalling Eq. (\ref{alpha}) and the boundary conditions of Eq. (\ref{matchingC}),
we get 
\begin{equation}
\frac{A^{2}+B^{2}}{2\kappa}-\frac{\kappa\lambda_{E}}{E_{0}}(A^{2}-B^{2})=1.\label{a-b}
\end{equation}
In terms of $A$ we can write 
\begin{equation}
A=(1-\lambda_{E})\sqrt{\frac{(1+\lambda_{E}^{2})\mu}{(1+\lambda_{E}^{2})^{3}+4\lambda_{E}^{2}\mu^{2}/E_{0}}}.
\end{equation}
After solving Eq. (\ref{a-b}) for $B$, we finally obtain the normalized
wave function, 
\begin{eqnarray}
\psi(x) & = & \sqrt{\frac{(1+\lambda_{E}^{2})\mu}{(1+\lambda_{E}^{2})^{3}+4\lambda_{E}^{2}\mu^{2}/E_{0}}}\nonumber \\
 &  & \times\left[\left(1-\lambda_{E}\right)e^{\kappa x}H(-x)+\left(1+\lambda_{E}\right)e^{-\kappa x}H(x)\right],
\end{eqnarray}
which, when we make $E_{0}\rightarrow\infty$,  recovers  the result found in Ref. \cite{Gadella-2009}, namely
\begin{equation}
\psi(x)=\frac{\sqrt{\mu}}{1+\lambda_{0}^{2}}\left[\left(1-\lambda_{0}\right)e^{\kappa x}H(-x)+\left(1+\lambda_{0}\right)e^{-\kappa x}H(x)\right].
\end{equation}

\begin{figure}[t]
\begin{centering}
\includegraphics[width=0.9\columnwidth]{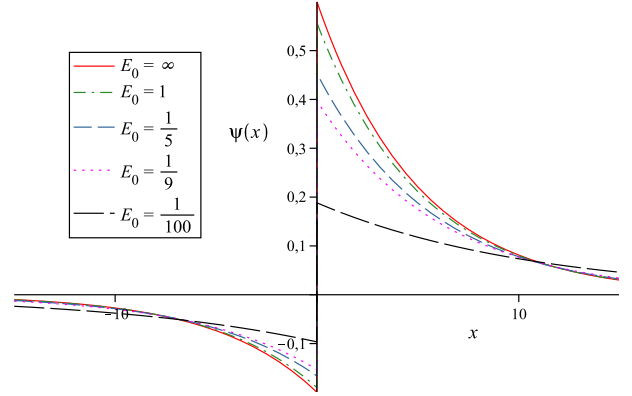} 
\par\end{centering}

\caption{Wave function $\psi (x)$ with $\mu=1$, $\lambda_{0}=2$ and several values for
$E_{0}$.\label{Wave-function}}
\end{figure}

In Fig. \ref{Wave-function} we can observe that for smaller values
of $E_{0}$ the wave function also decreases its amplitude. In the
limit $E_{0}\rightarrow0$, the Hamiltonian (\ref{Hamil}) tends to
a delta distribution, leading to a continuous wave function, as expected.

In Ref. \cite{Gadella-2009} it is pointed out that for $\lambda_{0}=\pm1$
the wave function exists only on one half of the $x-axis$. Here we
have the same situation, but in our model this happens for $\lambda_{E}=\pm1$.
Nevertheless, in order to maintain a fixed value for $\lambda_{E}$,
we need to correlate the parameters $\lambda_{0}$, $\mu$ and $E_{0}$.
For example, if we set $\lambda_{E}=1$, Eq. (\ref{eq:Energy}) furnishes
\begin{equation}
\lambda_{0}\exp\left(\frac{\mu^{2}}{8E_{0}}\right)=1.\label{relatexp}
\end{equation}
What is not seen when the interaction is energy-independent, i.e.
the fixation of one parameter does not imply any relation to another.

\begin{figure}[t]
\begin{centering}
\includegraphics[width=0.9\columnwidth]{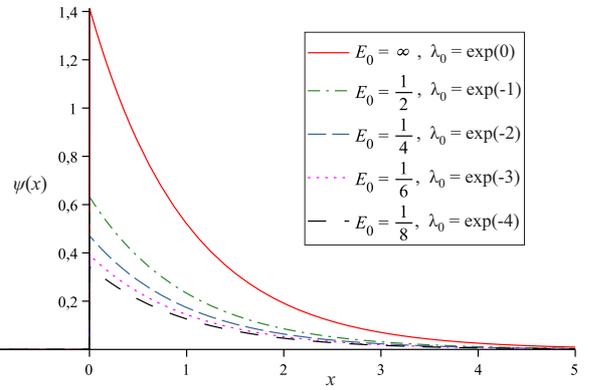} 
\par\end{centering}

\caption{Wave function of the bound state for different values of $E_{0}$
and $\lambda_{0}$ satisfying the relation $\lambda_{0}\exp[\mu^{2}/(8E_{0})]=1$
( $\mu=2$).\label{umlado}}
\end{figure}

On Figure \ref{umlado} its is shown the behavior of the wave function
for several values of the parameters satisfying Eq. (\ref{relatexp}),
again we see that for lower values of $E_{0}$ the wave function smoothly
decreases its amplitude.

%%%%%%%%%%%%%%%%%%%%%%%%%%%%%%%%%%%%%%%%%%%%%%%%%%%%%%%%%%%%%%%%%%%%%%%%%%%%%%%%%%%%%
\section{Final Remarks\label{sec:Final-Remarks}}

The interaction described by a $\delta$ distribution [Eq. (\ref{pure-delta})] 
naturally leads to full transmission in the limit
of high-energy incident particles [Eq. (\ref{full0})].
The introduction of a $\delta^{\prime}$
interaction term {[Eq. (\ref{primo})]} removes this characteristic
from the model [Eq. (\ref{eq:full1})]. 
In the present paper, considering the model shown
in Ref. \cite{Coutinho-JPA}, we proposed a solution for this problem
by introducing, for the $\delta^{\prime}$ term [Eq. (\ref{nosso-modelo})], 
an energy-dependent coupling parameter that goes to zero
for high energies [Eq. (\ref{lim-F})], so that the full transmission limit is achieved.

Considering the Hamiltonian (\ref{Hamil}) with (\ref{f(E)}), we calculated the
scattering coefficients and demonstrated that our model produces the
required limit for high energy incident particles [Eq. (\ref{limite-ok})]. 
This can be seen in Figure \ref{fig:Plot-of-the},
which exhibits the curves for some values of $E_{0}$, revealing that,
except for $E_{0}\rightarrow\infty$ (this limit recovers the case investigated in the literature \cite{Gadella-2009}),
the transmission tends to one for high energies.

In a similar way to Ref. \cite{Gadella-2009}, 
considering the Hamiltonian (\ref{Hamil}) with (\ref{f(E)}), we obtained a
more general relation for the bound state energy using the distribution
theory for discontinuous functions. Figure \ref{fig:energy} indicates
that the parameter $E_{0}$ is responsible for decreasing the energy
amplitude ($\lambda_{0}\neq0$), for which the effects of $E_{0}$
are more noticeable for larger values of $\lambda_{0}$ and $\mu$.
We also computed the bounded wave function 
shown in Figure \ref{Wave-function}, observing that the wave amplitude diminishes for smaller values
of $E_{0}$. A normalization scheme more appropriate for energy-dependent
potentials was carried out, as suggested in Ref. \cite{Lombard-04},
demonstrating that the amplitude of the wave function is not obtained
merely by making $\lambda\rightarrow\lambda_{0}e^{-E/E_{0}}$ in the
model of Ref. \cite{Gadella-2009}.

We found that (\ref{Hamil}) and (\ref{f(E)}) lead to 
a set of possibilities for the wave function to be different from zero only in half
of the $x$-axis [Eq. (\ref{relatexp})]. In the model discussed in
Ref. \cite{Gadella-2009}, these possibilities are limited
to the cases of Hamiltonians with the $\delta^{\prime}$-coefficient equal to $\pm1$. 

%a particle has the probability
%of living on one half of the axis in a set of possible ways, having
%
%the same condition, however, since we are dealing with energy-dependent
%parameters, imposing fixed values for this coefficient creates a relation
%between the parameters of the theory, as can be seen in Eq. (\ref{relatexp}).
%Physically, this means that, oppositely to Ref. \cite{Gadella-2009}
%(for which $E_{0}\rightarrow\infty$), a particle has the probability
%of living on one half of the axis in a set of possible ways, having
%in mind that we have chosen the appropriate parameters for this to
%happen, as can be seen in Figure \ref{umlado}.

Finally, we remark that, although the main problem motivating the 
consideration of the modified $\delta^{\prime}$ term in the $\delta-\delta^{\prime}$ Hamiltonian [ Eq. (\ref{nosso-modelo})]
was to describe potentials full transparent for high energy incident particles, 
this consideration also offers an additional degree of freedom stored in the choice of the function 
$F(E)$ for modeling the properties of transparency.
Intending to provide a similar degree of freedom 
to the $\delta$ term, we can extend the Hamiltonian in Eq. (\ref{nosso-modelo}) to
\begin{equation}
H=-\frac{1}{2}\frac{\partial^{2}}{\partial x^{2}}-\delta(x)\hat{G}+
\delta^{\prime}(x)\hat{F}
\label{nosso-modelo-2}
\end{equation}
where $\hat{G}$ is an operator described in a similar way as done for $\hat{F}$
in Eqs. (\ref{def-F}) and (\ref{lim-F}).
Mapping $\hat{F}\rightarrow 0$ in Eq. (\ref{nosso-modelo-2}), 
we get a modified pure $\delta$ model which extends the model (\ref{pure-delta})
considered in the literature \cite{Cohen}.

In summary, the model we have developed here has the feature of leading
with two important branches of quantum mechanics, which are: point
interactions and energy-dependent potentials. A correlation between
these models was made as a way to create
a more realistic model, compared to the cases when the potential is
energy-independent. Since we noticed that the transmission is not
full at high energies, but instead a constant dependent on the $\delta^{\prime}(x)$
coupling term, we have made such parameter a function that decreases
with the energy, obtaining the proper transmission. Remarkably, the
aforementioned considerations had several impacts on the bound state
solution, generating a wider class of physical situations.

\section*{Acknowledgments}

This work was partially
supported by CAPES and CNPq Brazilian agencies.

\end{document}